# Steady state quantum mechanics of thermally relaxing systems

Dvira Segal and Abraham Nitzan

School of Chemistry, The Sackler Faculty of Science, Tel Aviv University,

Tel Aviv, 69978, Israel

## Abstract

A theoretical description of quantum mechanical steady states is developed. Applications for simple quantum mechanical systems described in terms of coupled level structures yield a formulation equivalent to time independent scattering theory. Applications to steady states of thermally relaxing systems leads to time independent scattering theory in Liouville space that is equivalent to the tetradic Green's function formalism. It provides however a direct route to derive particular forms of the Liouville equation applicable in steady state situations. The theory is applied to study the conduction properties in the super-exchange model of a metal-molecule-metal contact weakly coupled to the thermal environment. The energy resolved temperature dependent transmission probability, as well as its coherent (tunneling) and incoherent (activated) parts, are calculated using the Redfield approximation. These components depend differently on the energy gap (or barrier), on the temperature and on the bridge length. The coherent component is most important at low temperatures, large energy gaps and small chain lengths. The incoherent component dominates in the opposite limits. The integrated transmission provides a generalization of the Landauer conduction formula for small junctions in the presence of thermal relaxation.

## 1. Introduction

Consider a system of classical rate equations. A set of variables C satisfies kinetic equations of the general form

$$\dot{\mathbf{C}} = \mathbf{F}(\mathbf{C}) \tag{1}$$

where the rates F are functions of the variables C. In many situations the set C is a continuous field (position dependent variable) and F contains differential and/or integral operators. As t→∞ the system will approach equilibrium if the rate laws in (1) satisfy detailed balance. When the boundary conditions imposed on the system are not compatible with equilibrium the set (1) may approach a non-equilibrium steady state (this will always be the case if the rates F are linear in the variables C) in which a constant current is passing through the system. The steady state is described by the set of equations

$$\mathbf{F}(\mathbf{C}) = 0 \tag{2}$$

together with boundary conditions (e.g. the values of some of the variables) that will characterize the non-equilibrium nature of the steady state. For example, the set of equations used to define the Lindeman mechanism in chemical kinetics[1,2]

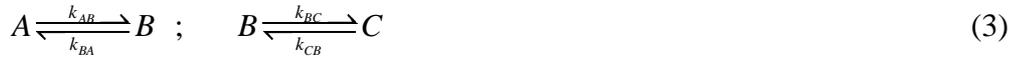

$$A \underset{k_{BA}}{\overset{k_{AB}}{\rightleftarrows}} B \ ; \quad B \underset{k_{CB}}{\overset{k_{BC}}{\rightleftarrows}} C \tag{3}$$

is often analyzed under the boundary conditions $C_A$=constant and $C_C$=0, where $C_i$ is the concentration of species *i*. Under these conditions a constant flux, $k_{AB}k_{BC}C_A/(k_{BC}+k_{BA})$ is passing through the system. In analogy, a non-equilibrium steady state of a diffusion process described by $\partial C(\mathbf{r},t)/\partial t = D\nabla^2 C(\mathbf{r},t)$ may be characterized by given constant values of $C(\mathbf{r},t)$ on opposite ends of the system. The steady state diffusion flux is $D\nabla C_{ss}(\mathbf{r})$, where $C_{ss}(\mathbf{r})$ is the solution to $D\nabla^2 C = 0$ under the given boundary conditions.

Quantum mechanical problems are rarely treated in a similar way. Boundary value problems are encountered mostly in the solution of the time independent Scrödinger equation, aimed at evaluating eigenstates of the system's Hamiltonian, which in themselves have no dynamical contents. Time dependent processes are treated as initial value problems. A prominent exception is the formulation of time independent scattering theory where the resulting wavefunctions can be interpreted as steady state solutions of a process characterized by a constant incoming flux. Scattering theory however is formulated in a particular framework in which incoming and outgoing waves become flux-carrying eigenstates of the free particle Hamiltonian far from an interaction zone. A more general formulation of steady state quantum mechanics can use different basis sets, e.g. states that by themselves do not carry fluxes. We have recently shown[3] that such a formulation can lead to standard scattering theory results as well as other results usually obtained from solving initial value quantum mechanical problems. In another recent paper[4] we have used a similar approach within a density matrix formalism for the analysis of thermal effects in electron transfer problems. Yet another important example where this approach is useful (and where an early version was used[5]) is light scattering, where thermal effects near resonance may increase the yield of fluorescence at the expense of Raman scattering. The general principles regarding this approach are summarized in Appendix A.

The purpose of the present paper is to point out, and to elucidate some subtle points in the application of the same approach within the quantum dynamical density matrix formalism. Our motivation is as in Ref. [4]: to develop a formalism for the description of steady state currents in metal-molecule-metal junctions, in particular in the presence of thermal interactions, and thereby to evaluate the conduction properties of such junctions. This paper focuses on technical aspects associated with the application of this technique. In a subsequent paper[6] we will use this approach to study the heat dissipation in a model for steady state charge transfer through a molecular bridge.

## 2. Simple examples

To set the stage for the more complex problems discussed below we review in this section the application of the steady state technique to simple problems involving the decay of an initially prepared ("doorway") state interacting with a continuum. Figure 1 shows the standard model for this process: The Hamiltonian is expressed in terms of an orthonormal basis

$$H = E_0 |0><0| + \sum_j E_j |j><j| + \sum_j \left( V_{0j} |0><j| + V_{j0} |j><0| \right) \qquad (4)$$

where the states $\{j\}$ constitute a quasi-continuous manifold, characterized by its density $\rho_J(E) = \sum_j \delta(E - E_j)$, which extends into energies far below and above $E_0$. The coupling matrix elements $V_{ij} = V_{ji}^*$ are assumed to depend weakly on $E_j$ and to vanish near the edges of the $\{j\}$ spectrum. Under these (and some other well known) conditions an initial state $\psi(t=0) = |0>$ evolves in time according to $<\psi(0)|\psi(t)> = \exp\left(-i\tilde{E}_0 t - (1/2)\Gamma_{0J}(E_0)t\right)$ where $\tilde{E}_0 = E_0 + \Lambda_{0J}(E_0)$ and where $\Lambda_{0J}(E)$ and $\Gamma_{0J}(E)$ are the real and (-)twice the imaginary parts of the function

$$\Sigma_{0J}(E) = \lim_{\eta \to 0} \sum_j \left(|V_{0j}|^2 / (E - E_j + i\eta)\right) \tag{5}$$

In particular,

$$\Gamma_{0J}(E_0) = 2\pi \sum_j |V_{0j}|^2 \delta(E_0 - E_j) \equiv 2\pi \left[|V_{0J}|^2 \rho_J\right]_{E=E_0} \tag{6}$$

($|V_{0J}|$ is defined by this relation), is the well known golden rule expression for the decay rate of an initially prepared state coupled to a continuum, and is the main result of this model. Here and in the rest of the paper we have used capital characters to denote manifolds of states, while the corresponding lower case font denotes individual states.

In order to obtain this result from a steady state formulation we start from the equations of motion for the coefficients c of the expansion of a general solution of the time dependent Scrödinger equation for this model, $\Psi(t) = c_0(t)|0> + \sum_j c_j(t)|j>$:

$$\dot{c}_0 = -iE_0 c_0 - i\sum_j V_{0,j} c_j \quad ; \quad \dot{c}_j = -iE_j c_j - iV_{j,0} c_0. \tag{7}$$

Consider the steady state obtained if the state $|0>$ was *forced to evolve as if the coupling to the continuum did not exist* while each of the states of the manifold $\{j\}$ is assigned a small damping term $\eta/2$. This damping is an absorbing boundary condition that can be taken zero at the end of the calculation. The corresponding equations are

$$c_0(t) = C_0 e^{-iE_0 t} \tag{8}$$

$$\dot{c}_j = -iE_j c_j - iV_{j,0} c_0 - \frac{\eta}{2} c_j \tag{9}$$

$c_0(t)$ given by (8) is now a driving term in Eq. (9) for $c_j$. The latter yields, at long time $c_j(t) = C_j e^{-iE_0 t}$, with $C_j$ given by

$$C_j = \frac{V_{j,0} C_0}{(E_0 - E_j) + i\eta/2} \tag{10}$$

The total flux through the systems is

$$I = \lim_{\eta \to 0} \left(\eta \sum_j |C_j|^2\right) = |C_0|^2 \Gamma_{0J} \tag{11}$$

implying a rate, $I/|C_0|^2$, equal to $\Gamma_{0J}$. To reiterate, the procedure just described replaces the original Shrödinger equation by an equation that incorporates two boundary conditions: The driving term, Eq. (8), corresponds to a constant incoming flux while the absorbing boundary terms imposed on Eqs. (9) cause the (linear) system to approach a steady state as t→∞. The fact that an analysis of the relationship between given non-equilibrium boundary conditions and the steady state flux sustained by them can yield information on rates is well known in kinetic theory. However, the *steady state rate* and the rate observed in a transient measurement are not always the same; for a discussion of this point see Ref. [4].

Next consider another example: a model for resonance scattering. Figure 2 depicts a typical potential model for this problem, together with a corresponding energy level structure. The potential scattering problem, fig. 2a, corresponds to what we normally refer to as resonance tunneling: A particle with energy $E_0$ approaches the double barrier structure from the left, and the transmission and reflection probabilities are evaluated as functions of $E_0$. Fig. 2b provides an approximation to this model, based on the assumption that the scattering process is dominated by the interaction of the incident and scattered waves with the single resonant level $|1>$ in the well between the barriers. Fig. 2c is a generalization to the case with many intermediate states (or wells) that will be discussed later. Fig. 2b is also a model for absorption and resonance scattering of light: Here level $|0>$ represents a dressed state, e.g. the molecular ground state $|g>$ dressed by a photon so that $E_0 = E_g + \hbar\omega$, while state $|1>$ is the molecular excited state without the photon. The corresponding Hamiltonian is

$$H = H_0 + V \tag{12}$$

$$H_0 = E_0 |0><0| + E_1 |1><1| + \sum_{l \neq 0} E_l |l><l| + \sum_r E_r |r><r| \tag{13}$$

$$V = \sum_{l \neq 0} \left(V_{l,1} |l><1| + V_{1,l} |1><l|\right) + \left(V_{0,1} |0><1| + V_{1,0} |1><0|\right) + \sum_r \left(V_{r,1} |r><1| + V_{1,r} |1><r|\right) \tag{14}$$

Where $\{r\}$ and $\{l\}$ now denote the right and left continuous manifolds. The incident state $|0>$ belongs to the $\{l\}$ manifold but should be treated separately as discussed below. In collision theory $|0>$ represent an incoming wave while the other $\{l\}$ states and the $\{r\}$ states correspond to outgoing waves. These incoming and outgoing waves carry momentum, but the formalism described here can use other representations, including one in which these states are standing waves.[3]

A general solution of the Schrödinger equation based on the Hamiltonian (12)-(14) takes the form $\Psi(t) = \sum_j c_j(t)|j>$ ; $j = 0,1,\{l\},\{r\}$, where the coefficients **c** satisfy

$$\dot{c}_0 = -iE_0 c_0 - iV_{0,1} c_1 \tag{15}$$

$$\dot{c}_1 = -iE_1 c_1 - iV_{1,0} c_0 - i\sum_l V_{1,l} c_l - i\sum_r V_{1,r} c_r \tag{16}$$

$$\dot{c}_l = -iE_l c_l - iV_{l,1} c_1 \tag{17}$$

$$\dot{c}_r = -iE_r c_r - iV_{r,1} c_1 \tag{18}$$

In analogy to Eqs. (8) and (9) we study instead a set that would lead to a steady state at long time

$$\dot{c}_0 = -iE_0 c_0 \tag{19}$$

$$\dot{c}_1 = -iE_1 c_1 - iV_{1,0} c_0 - i\sum_l V_{1,l} c_l - i\sum_r V_{1,r} c_r \tag{20}$$

$$\dot{c}_l = -iE_l c_l - iV_{l,1} c_1 - (\eta/2) c_l \tag{21}$$

$$\dot{c}_r = -iE_r c_r - iV_{r,1} c_1 - (\eta/2) c_r \tag{22}$$

Note that while Eq. (15) is just one of equations (17) (since the incident state belongs to the manifold $\{\ell\}$), in Eq. (19) this state is given a special status as the one that drives the system (see also discussion at the end of this section). As t→∞ all coefficients **c** oscillate with this driving frequency

$$c_j(t) = C_j e^{-iE_0 t} \quad ; \quad j = 0,1,l,r \tag{23}$$

using this in (20)-(22) leads to

$$C_r = \frac{-V_{r,1} C_1}{E_r - E_0 - i\eta/2} \quad \Rightarrow \quad -i\sum_r V_{1,r} C_r \equiv -i\Sigma_{1R}(E_0) C_1 \tag{24}$$

$$\Sigma_{1R}(E_0) = \Lambda_{1R}(E_0) - (1/2) i \Gamma_{1R}(E_0)$$

and similar equation for $C_\ell$ and $-i\sum_l V_{1,l} C_l$. Using these results in (20) yields

$$C_1 = \frac{V_{1,0} C_0}{E_0 - \tilde{E}_1 + (i/2)\Gamma_1(E_0)} \tag{25}$$

where $\Gamma_1(E_0) = \Gamma_{1L}(E_0) + \Gamma_{1R}(E_0)$ and where $\tilde{E}_1 = E_1 + \Lambda_{1R}(E_0) + \Lambda_{1L}(E_0)$ is the shifted resonance energy. Using (25) in the expression for $C_r$ in (24) we get

$$|c_r|^2 = |C_r|^2 = \frac{|V_{r,1}|^2}{(E_r - E_0)^2 + (\eta/2)^2} \frac{|V_{1,0}|^2 |c_0|^2}{(\tilde{E}_1 - E_0)^2 + (\Gamma_1(E_0)/2)^2} \tag{26}$$

for the population of the $r$ state with energy $E_r$ in the right continuous manifold. The steady-state flux to the right out of this level is given by $\eta |c_r|^2$ and the corresponding rate $k_{0 \to r}$ is obtained by dividing by $|c_0|^2$. In the limit $\eta \to 0$ this becomes

$$k_{0 \to r} = 2\pi |V_{1,r}|^2 \, \delta(E_r - E_0) \frac{|V_{1,0}|^2}{(\tilde{E}_1 - E_0)^2 + (\Gamma_1(E_0)/2)^2} \tag{27}$$

while the total rate to the right is

$$k_{0 \to R} = \sum_r k_r = \Gamma_{1R}(E_0) \frac{|V_{1,0}|^2}{(\tilde{E}_1 - E_0)^2 + (\Gamma_1(E_0)/2)^2} \tag{28}$$

Similar expressions (e.g. $\Gamma_{1L}$ replaces $\Gamma_{1R}$ in (28)) are obtained for the leftward flux. The energy conservation implied by the delta-function in Eq. (27) is to be expected in the present case where no thermal dissipation is taking place. These expressions may be rewritten in symmetric forms more closely related to scattering theory. The rate into the right manifold per unit final energy is[7]

$$k_{0 \to R}(E) \equiv \sum_r k_{0 \to r} \delta(E - E_r) = \Gamma_{1R}(E) \frac{|V_{1,0}|^2}{(\tilde{E}_1 - E_0)^2 + (\Gamma_1(E_0)/2)^2} \delta(E - E_0) \tag{29}$$

on the other hand, this rate is related to the transmission coefficient $\mathcal{T}$ by

$$k_{0 \to R}(E) = \frac{q_0}{mL} \mathcal{T}(E_0, E) \tag{30}$$

where $q_0$ is the incident carrier momentum, m - the carrier mass and where $L$ is the normalization length. In terms of the 1-dimensional density of states(8) $\rho_L(E_0) = Lm/(2\pi q_0)$ this implies

$$\mathcal{T}_{el}(E_0, E) = 2\pi \rho_L(E_0) k_{0 \to R}(E) = \frac{\Gamma_{1L}(E_0)\Gamma_{1R}(E_0)}{(\tilde{E}_1 - E_0)^2 + (\Gamma_1(E_0)/2)^2} \delta(E - E_0) \tag{31}$$

(The subscript "*el*" is used to denote the elastic character of the transmission process) The expression multiplying the delta function corresponds to the usual definition of transmission coefficient at energy $E_0$ is

$$\mathcal{T}_{el}(E_0) = \int dE \, \mathcal{T}_{el}(E_0, E) = \frac{\Gamma_{1L}(E_0)\Gamma_{1R}(E_0)}{(\tilde{E}_1 - E_0)^2 + (\Gamma_1(E_0)/2)^2} \tag{32}$$

Note that all the damping and shift terms in Eqs. (27)-(32) are evaluated at $E=E_0$. A common approximation to the solution of Eqs. (19)-(22) is obtained by accounting for the interaction between level $|1\rangle$ and the $\{l\}$ and $\{r\}$ manifolds by replacing the sums in Eq. (16) by appropriate damping and shift terms computed at $E=E_1$. Eq. (20) is then replaced by

$$\dot{c}_1 = -i\tilde{E}_1 c_1 - iV_{1,0}c_0 - (1/2)[\Gamma_{1L}(E_1) + \Gamma_{1R}(E_1)]c_1 \qquad (33)$$

When used in conjunction with $c_1 = C_1 e^{-iE_0 t}$ (c.f. Eq. (23) this leads to an equation similar to (25), except that $\Gamma_1(E_0)$ and $\Lambda_1(E_0)$ are replaced by $\Gamma_1(E_1)$ and $\Lambda_1(E_1)$. This in turn leads to equations similar to (27)-(32) with similar substitutions. It is seen that this provides a good approximation only in resonant cases ($E_0 \approx E_1$) and in Markovian situations where $\Gamma$ and $\Lambda$ do not depend on $E$

The generalization of this result to the case of $N$ intermediate levels (Fig. 2c) follows the same steps. The Hamiltonian, Eqs. (12)-(14) is now

$$H = H_0 + V$$

$$H_0 = E_0 |0\rangle\langle 0| + \sum_{n=1}^{N} E_n |n\rangle\langle n| + \sum_{l \neq 0} E_l |l\rangle\langle l| + \sum_r E_r |r\rangle\langle r| \qquad (34)$$

$$V = \sum_{n=1}^{N} \sum_{n'=1}^{N} V_{n,n'} |n\rangle\langle n'| + \sum_{l \neq 0} \sum_{n=1}^{N} \left( V_{l,n} |l\rangle\langle n| + V_{n,l} |n\rangle\langle l| \right)$$
$$+ \sum_{n=1}^{N} \left( V_{0,n} |0\rangle\langle n| + V_{n,0} |n\rangle\langle 0| \right) + \sum_r \sum_{n=1}^{N} \left( V_{r,n} |r\rangle\langle n| + V_{n,r} |n\rangle\langle r| \right) \qquad (35)$$

where the coupling scheme is taken to be general, not necessarily nearest neighbor. The equations of motion equivalent to (20)-(22) are

$$\dot{c}_n = -iE_n c_n - iV_{n,0}c_0 - i\sum_{n' \neq n} V_{n,n'} c_{n'} - i\sum_{l \neq 0} V_{n,l} c_l - i\sum_r V_{n,r} c_r \qquad (36)$$

$$\dot{c}_j = -iE_j c_j - \sum_{n=1}^{N} iV_{j,n} c_n - (\eta/2)c_j \quad ; \qquad j = l, r \qquad (37)$$

Going to steady state, as before, and eliminating the $\{l\}$ and $\{r\}$ manifolds from Eq. (36) using the same procedure as in (24), we get ($C_n = c_n e^{iE_0 t}$ ; $E_{n,0} = E_n - E_0$)

$$0 = -iE_{n,0}C_n - iV_{n,0}C_0 - i\sum_{n' \neq n} V_{n,n'} C_{n'} - i\sum_{n'} \Sigma_{n,n'}(E_0) C_{n'} \qquad (38)$$

where the self-energy matrix $\Sigma$ is

$$\Sigma_{n,n'}(E_0) = \Sigma_{n,n'}^{(L)}(E_0) + \Sigma_{n,n'}^{(R)}(E_0)$$
$$\Sigma_{n,n'}^{(J)}(E) = \sum_j \frac{V_{n,j}V_{j,n'}}{E - E_j + i\eta/2} = \Lambda_{n,n'}^{(J)}(E) - \frac{1}{2}i\Gamma_{n,n'}^{(J)}(E) \quad ; \quad J = L, R \qquad (39)$$

Defining the vectors

$$\mathbf{C} = \begin{pmatrix} C_1 \\ \vdots \\ C_N \end{pmatrix} \quad ; \quad \mathbf{V}_0 = \begin{pmatrix} V_{1,0} \\ \vdots \\ V_{N,0} \end{pmatrix} \tag{40}$$

and the effective Hamiltonian matrix in the subspace of intermediate states 1...N

$$H_{n,n'}^{(N)} = E_n \delta_{n,n'} + V_{n,n'} + \Sigma_{n,n'} \tag{41}$$

equations (38) can be recast in the form

$$\left(E_0 - \mathbf{H}^{(N)}\right)\mathbf{C} = \mathbf{V}_0 C_0 \quad \Rightarrow \quad \mathbf{C} = \left(E_0 - \mathbf{H}^{(N)}\right)^{-1} \mathbf{V}_0 C_0 \tag{42}$$

Eq. (42) gives the steady state amplitudes for all intermediate states $\{n\}=1...N$. The steady state rate is obtained from $k_{0 \to r} = \eta |C_r|^2 / |C_0|^2$ where $C_r = c_r e^{iE_0 t}$ is obtained from the steady state version of (37). This leads to

$$k_{0 \to r} = \frac{\eta}{(E_r - E_0)^2 + (\eta/2)^2} \frac{1}{|C_0|^2} |\sum_n V_{r,n} C_n|^2 \xrightarrow{\eta \to 0}$$

$$2\pi \delta(E_r - E_0) \left| \sum_n \sum_{n'} V_{r,n} G_{n,n'}^N V_{n'0} \right|^2 \tag{43}$$

Repeating the steps that lead from Eq. (27) to (31) we now get

$$\mathcal{T}_{el}(E_0, E) = 2\pi \rho_L(E_0) \sum_r k_{0 \to r} \delta(E - E_r)$$
$$= \delta(E - E_0) \sum_n \left( G^{(N)} \Gamma^{(L)} G^{(N)+} \Gamma^{(R)} \right)_{n,n} = \delta(E - E_0) Tr_N \left( G^{(N)} \Gamma^{(L)} G^{(N)+} \Gamma^{(R)} \right) \tag{44}$$

So that the transmission coefficient is

$$\mathcal{T}_{el}(E_0) = Tr_N \left( G^{(N)}(E_0) \Gamma^{(L)}(E_0) G^{(N)+}(E_0) \Gamma^{(R)}(E_0) \right) \tag{45}$$

in agreement with results obtained from standard scattering theory. We note in passing that if the $\{l\}$ and $\{r\}$ manifolds correspond to metal electrodes on the two sides of a molecular constriction represented by the states $\{n\}$, the corresponding conduction at zero bias is given by the Landauer formula[9,10]

$$g = \frac{e^2}{\pi \hbar} \mathcal{T}_{el}(E_F) \tag{46}$$

where $E_F$ is the Fermi energy.

Up to this point our discussion may be regarded as a reformulation of scattering theory. This reformulation has the important attribute that it is not restricted to use wavefunctions that satisfy the usual incoming and outgoing boundary conditions. Rather, any chosen state may be taken to drive the system and the consequences of this driving may be studied. The need to reformulate scattering theory in this language arises from the nature of some applications, e.g. current in a metal-insulator-metal junction, where, in the weak coupling limit, representing a process in terms of 'left' and 'right' manifolds of standing wave states is natural.[11] This requires, as already noted, exercising some caution in distinguishing between the driving state and the manifold of states it belongs to. In the application described above, even though the state $|0>$ formally belongs to the manifold $\{l\}$, it has a special status as the state that drives the system into a non-equilibrium steady state. Any error due to double counting in sums over states of this manifold is negligible due to the huge number of such states.

## 3. Steady state quantum mechanics of thermally relaxing systems

The effect of thermal dephasing and relaxation on the dynamics of a quantum system may be studied using a suitable density matrix formalism. Here we focus on scattering processes in which the scattering particle interacts with the thermal environment of the target. Raman scattering from molecules in solution and electron tunneling in metal-insulator-metal junctions where the metal electrons are modeled as free particles, are example of such processes.

It is useful to see first how steady state phenomena are described within the density matrix framework in athermal situations. To this end consider again the system represented by the Hamiltonian (12)-(14). A set of dynamical equations equivalent to (15)-(18) may be written for elements of the density matrix, $\rho_{i,j}$. Formally they can be derived from (15)-(18) or from the Liouville equation

$$\frac{d\rho}{dt} = -i[H, \rho] \tag{47}$$

supplemented by the damping ($\eta$) terms. This leads to

$$\dot\rho_{0,0} = -iV_{0,1}\rho_{1,0} + iV_{1,0}\rho_{0,1} \tag{48}$$

$$\dot\rho_{0,1} = -iE_{0,1}\rho_{0,1} - iV_{0,1}(\rho_{1,1} - \rho_{0,0}) + i\sum_j V_{j,1}\rho_{0,j} \tag{49}$$

$$\dot\rho_{0,j} = -iE_{0,j}\rho_{0,j} - iV_{0,1}\rho_{1,j} + iV_{1,j}\rho_{0,1} - \eta/2\rho_{0,j} \tag{50}$$

$$\dot\rho_{1,1} = -iV_{1,0}\rho_{0,1} + iV_{0,1}\rho_{1,0} - i\sum_j V_{1,j}\rho_{j,1} + i\sum_j V_{j,1}\rho_{1,j} \tag{51}$$

$$\dot\rho_{1,j} = -iE_{1,j}\rho_{1,j} - iV_{1,0}\rho_{0,j} - i\sum_{j'} V_{1,j'}\rho_{j',j} + iV_{1,j}\rho_{1,1} - \eta/2\rho_{1,j} \tag{52}$$

$$\dot\rho_{j',j} = -iE_{j',j}\rho_{j',j} - iV_{j',1}\rho_{1,j} + iV_{1,j}\rho_{j',1} - \eta\rho_{j',j} \tag{53}$$

Where $E_{\alpha,\beta} = E_\alpha - E_\beta$. $j'$ and $j$ denote states from the $\{l\}$ or $\{r\}$ manifolds and the sums are over both manifolds. For any pair of indices $\alpha\beta$, the equation of motion for $\rho_{\beta,\alpha}$ is the complex conjugates of that for $\rho_{\alpha,\beta}$.

Within this Liouville formalism, what are the steady state equations that correspond to (19)-(22)? Recalling that the system is driven by the state $|0>$ and that at steady state all the amplitudes satisfy Eq. (23), imply that all $\rho_{I,j}$ are constants at steady state. This may suggests that a proper set of steady state equations is obtained from (48)-(53) by (a) replacing Eq. (48) by $\rho_{0,0}$=constant and by setting all $\dot\rho$ on the left side of Eqs. (49)-(53) to zero. The resulting set of equations indeed describes a quantum mechanical steady state driven by a maintained constant *population* in state $|0>$. This is however not equivalent to the steady state described by Eqs. (19)-(22), which is driven by the fixed amplitude and *phase* of state $|0>$ (c.f. Eq. (23)). This can be easily realized by using Eqs. (19)-(22) together with $\rho_{i,j} = c_i c_j^*$ to derive the following set of steady state equations

$$\rho_{00} = \text{constant} \tag{54}$$

$$\dot\rho_{0,1} = 0 = -iE_{0,1}\rho_{0,1} + iV_{0,1}\rho_{0,0} + i\sum_j V_{j,1}\rho_{0,j} \tag{55}$$

$$\dot\rho_{0,j} = 0 = -iE_{0,j}\rho_{0,j} + iV_{1,j}\rho_{0,1} - (\eta/2)\rho_{0,j} \tag{56}$$

$$\dot\rho_{1,1} = 0 = -iV_{1,0}\rho_{0,1} + iV_{0,1}\rho_{1,0} - i\sum_j V_{1,j}\rho_{j,1} + i\sum_j V_{j,1}\rho_{1,j} \tag{57}$$

$$\dot\rho_{1,j} = 0 = -iE_{1,j}\rho_{1,j} - iV_{1,0}\rho_{0,j} - i\sum_{j'} V_{1,j'}\rho_{j',j} + iV_{1,j}\rho_{1,1} - (\eta/2)\rho_{1,j} \tag{58}$$

$$\dot\rho_{j',j} = 0 = -iE_{j',j}\rho_{j',j} - iV_{j',1}\rho_{1,j} + iV_{1,j}\rho_{j',1} - \eta\rho_{j',j} \tag{59}$$

Comparing to Eqs. (49)-(53) reveals that two terms, $iV_{0,1}\rho_{1,1}$ from the RHS of Eq. (49) and $iV_{0,1}\rho_{1,j}$ from the RHS of Eq. (50), are absent in Eqs. (55) and (56). Since equations (54)-(59) are equivalent to Eqs. (19)-(22) they would lead to the same results, e.g., Eqs. (27) and (28) as above. See also Appendix B.

It may seem peculiar that the set of equations (48)-(53), a formally rigorous representation of the Liouville equation (47) has to be replaced by the set (54)-(59) in which particular terms are missing. We should keep in mind however that we are dealing with a reformulation of scattering theory where states $|0>$ and $\{|j>\}$ are normalized in an infinite volume. Therefore all matrix elements $V_{0,1}$ or $V_{1,j}$ scale like $\Omega^{-1/2}$, where $\Omega \to \infty$ is the normalization volume. Therefore, in evaluating transmission coefficients as in Eqs. (44)-(45) only terms of two kinds survive: either coupling of intermediate (bridge) states (here |1>) to the continuous manifolds that appear in damping and shift terms, e.g. $|V_{1,j}|^2 \rho_J$, or coupling of the driving state |0> to an intermediate state, that appear in the influx term $|V_{1,0}|^2 \rho(E_0)$. Both combinations are independent of $\Omega$. The terms discarded in Eqs. (54)-(59) are those that contribute terms like $\rho_L |V_{1,0}|^\xi ; \xi > 2$ that vanish in the limit $\Omega \to \infty$.[12]

Next consider a thermal system described by the Hamiltonian

$$\mathcal{H} = H + H_B + F \tag{60}$$

where $H$ is the Hamiltonian (12)-(14) that corresponds to the generic model for resonance scattering of Fig. 2b, $H_B$ is thermal bath Hamiltonian and $F$ – a system-bath coupling, here taken to be coupled diagonally to the resonance state $|1>$

$$F = F_{11} |1><1|. \tag{61}$$

The exact form of $F$ is not important, but in the present discussion we will assume that this coupling to the thermal environment is weak.(13) This coupling is characterized by its correlation function, whose Fourier transform satisfies the detailed balance relation

$$\int_{-\infty}^{\infty} dt e^{i\omega t} <F_{11}(t)F_{11}(0)> = e^{\beta \hbar \omega} \int_{-\infty}^{\infty} dt e^{i\omega t} <F_{11}(0)F_{11}(t)> \quad ; \quad \beta = (k_B T)^{-1} \tag{62}$$

where T is the temperature and $k_B$ – the Boltzmann constant. For specificity we will sometime use

$$<F_{11}(t)F_{11}(0)> = \frac{\kappa}{2\tau_c} \exp(-|t|/\tau_c) \tag{63}$$

which becomes $\kappa \delta(t)$ in the Markovian, $\tau_c \to 0$, limit.

We write the Liouville equation $\dot{\rho} = -i[\mathcal{H}, \rho]$ in the form

$$\dot{\tilde{\rho}} = -i[H, \tilde{\rho}] - i[\tilde{F}, \tilde{\rho}] \tag{64}$$

where we have applied the transformation $\rho \to \tilde{\rho} \equiv e^{iH_B t} \rho e^{-iH_B t}$; $F \to \tilde{F} \equiv e^{iH_B t} F e^{-iH_B t}$, and where the two terms on the RHS of (64) correspond to 'deterministic' and 'thermal' contributions. In what follows we omit the tilde sign above the operators, keeping in mind that the following equations are written for the transformed operators. Our ultimate goal is to obtain the evolution of the reduced system's density matrix $\sigma = Tr_B \rho$. In other words, we want to find the steady state in the σ subspace that is determined by the same boundary conditions as in Eqs. (19)-(22) that led to Eqs. (54)-(59) in the athermal case.

It is easy to see that in the corresponding steady state equations the deterministic part of Eq. (64) leads again to Eqs. (54)-(59). These should be supplemented by terms arising from the interaction with the thermal bath, leading to

$$\rho_{00} = \text{constant} \tag{65}$$

$$\dot{\rho}_{0,1} = 0 = -iE_{0,1}\rho_{0,1} + iV_{0,1}\rho_{0,0} + i\sum_j V_{j,1}\rho_{0,j} - i[F,\rho]_{0,1} \tag{66}$$

$$\dot{\rho}_{0,j} = 0 = -iE_{0,j}\rho_{0,j} + iV_{1,j}\rho_{0,1} - (\eta/2)\rho_{0,j} - i[F,\rho]_{0,j} \tag{67}$$

$$\dot{\rho}_{1,1} = 0 = -iV_{1,0}\rho_{0,1} + iV_{0,1}\rho_{1,0} - i\sum_j V_{1,j}\rho_{j,1} + i\sum_j V_{j,1}\rho_{1,j} - i[F,\rho]_{1,1} \tag{68}$$

$$\dot{\rho}_{1,j} = 0 = -iE_{1,j}\rho_{1,j} - iV_{1,0}\rho_{0,j} - i\sum_{j'} V_{1,j'}\rho_{j',j} + iV_{1,j}\rho_{1,1} - (\eta/2)\rho_{1,j} - i[F,\rho]_{1,j} \tag{69}$$

$$\dot{\rho}_{j',j} = 0 = -iE_{j',j}\rho_{j',j} - iV_{j',1}\rho_{1,j} + iV_{1,j}\rho_{j',1} - \eta\rho_{j',j} \tag{70}$$

where, as before, the index $j$ corresponds to both the left and the right manifolds.

Next we make a simplifying approximation by assuming that thermal interactions can be disregarded in the evolution of matrix elements of $\rho$ that involve the continuous manifolds $\{j\}$. This implies that the commutators involving $F$ in Eqs. (67) and (69) are neglected. (The absence of such a term in (70) is a consequence of the form (61) of $F$). The rational for making this approximation is based on the expectation that because our thermal interactions are localized in the subspace of intermediate (bridge) states (here |1>), disregarding them in Eqs. (67) and (69) should not affect the dynamics in the continuum, while the effect on the bridge dynamics should be weak in the weak coupling limit used below (see also Sect. 5).

This in turn also implies that the procedure for replacing the sum over the {j} states (i.e. over the left and right manifolds) by damping terms can be done as if the thermal interactions were absent. This is an important technical detail, because it makes it possible to carry out this reduction procedure in the amplitude representation, starting from Eqs. (19)-(22), then use the reduced amplitude equations to evaluate the corresponding equations for the density matrix; see Appendix B. The resulting steady-state equations are

$$\dot{\rho}_{0,1} = 0 = -i\left(E_0 - \tilde{E}_1\right)\rho_{0,1} + iV_{0,1}\rho_{0,0} - \frac{1}{2}\Gamma_1\rho_{0,1} - i[F,\rho]_{0,1} \tag{71}$$

$$\dot{\rho}_{1,1} = 0 = -iV_{1,0}\rho_{0,1} + iV_{0,1}\rho_{1,0} - \Gamma_1\rho_{11} - i[F,\rho]_{1,1} \tag{72}$$

$$\dot{\rho}_{0,j} = 0 = -i(E_0 - E_j)\rho_{0,j} + iV_{1,j}\rho_{0,1} - (\eta/2)\rho_{0,j} \tag{73}$$

$$\dot{\rho}_{1,j} = 0 = -i(\tilde{E}_1 - E_j)\rho_{1,j} - iV_{1,0}\rho_{0,j} + iV_{1,j}\rho_{1,1} - \frac{1}{2}\Gamma_1\rho_{1,j} \tag{74}$$

$$\dot{\rho}_{j',j} = 0 = -i(E_{j'} - E_j)\rho_{j',j} - iV_{j',1}\rho_{1,j} + iV_{1,j}\rho_{j',1} - \eta\rho_{j',j} \tag{75}$$

where $\Gamma_1 = \Gamma_1(E_0)$ and where $\tilde{E}_1 = E_1 + \Lambda_1(E_0)$ are defined above and in Appendix B. In Eq. (74) we took $\Gamma + \eta \to \Gamma$.

The following points are notable: First, in the absence of thermal interactions, i.e. when the commutators involving $F$ are absent in Eqs. (71) and (72), Eqs. (71)-(75) lead to (see Appendix B) the steady state result

$$\eta\rho_{j,j} = 2\pi |V_{1,j}|^2 \delta(E_j - E_0) \frac{|V_{1,0}|^2 \rho_{0,0}}{(\tilde{E}_1 - E_0)^2 + (\Gamma_1/2)^2} \tag{76}$$

This will yield, e.g., Eq. (27) if applied to $j \in R$ (i.e., a state of the right manifold).

Secondly, under our approximations, the two equations (71) and (72) that, together with the boundary condition $\rho_{00}$=constant, describe a steady state in a damped and thermally relaxing two-level system, can be solved independently from Eqs. (73)-(75).

Third, the latter equations can be used to obtain a complete description of the scattering process: as before $\eta\rho_{j,j}$, which depends on the incident energy $E_0$, the resonance energy $E_1$ and the scattered energy $E_j$, is the steady state flux out of, hence also into, the final state $j$. This is a transmission flux for $j \in \{r\}$ and a reflection flux for $j \in \{l\}$. In the athermal situation (Eq. (76)) $\eta\rho_{jj} \sim \delta(E_0 - E_j)$. This is no longer true in the thermal case.

For simplicity we will disregard in what follows the energy dependence of the functions $\Gamma_1(E)$ and $D_1(E)$. Also, for simplicity of notation we will disregard the tilde above $E_1$, keeping in mind that $E_1$ represents the shifted resonance energy. Consider first Eqs. (73)-(75). Since these equations do not involve interaction with the heat bath, taking a trace over the bath states simply amounts to replacing $\rho$ by $\sigma$ everywhere in these equations. We focus on transitions into the 'right' manifold (e.g. in a metal-insulator-metal junction we study transfer from a particular level $|0\rangle$ on the metal on the left to the manifold of levels on the right). Eliminating $\sigma_{0r}$ using Eq. (73) yields

$$\eta\sigma_{r,r} = 2\,\text{Im}(V_{r,1}\sigma_{1,r}) \tag{77}$$

$$\sigma_{1,r} = \frac{V_{1,0}V_{1,r}}{E_{0,1} + i\Gamma_1/2} \frac{1}{E_{0,r} - i\eta/2} \sigma_{0,1} + \frac{V_{1,r}}{E_{1,r} - i(\Gamma_1 + \eta)/2} \left[ \sigma_{1,1} - \frac{V_{1,0}\sigma_{0,1}}{E_{0,1} + i\Gamma_1/2} \right] \quad (78)$$

where $E_{n,m} = E_n - E_m$.

To obtain $\eta \sigma_{r,r}$ we need to get $\sigma_{0,1}$ and $\sigma_{1,1}$ from Eqs. (71) and (72). Before imposing $\dot{\rho} = 0$ these equations describe the time evolution of a damped two-level system interacting with a heat bath; the corresponding Liouville equation is

$$\dot{\rho} = -i[H_0 + V, \rho] - i[F, \rho] - \frac{\Gamma_1}{2} \{|1><1|, \rho\} \quad (79)$$

where from here on we use $H_0$ to denote the zero order Hamiltonian in the 'system' subspace

$$H_0 = E_0 |0><0| + E_1 |1><1| \quad (80)$$

and

$$V = V_{0,1} |0><1| + V_{1,0} |1><0| \quad (81)$$

$\rho$ is the density operator in the system-bath space and $[,]$ and $\{,\}$ denote commutator and anticommutator respectively. To obtain the time evolution in the system subspace we follow the procedure of Ref.[4], which relies on the Redfield approximation.[14-16] This implies the assumption of weak coupling between the system and its thermal environment. As discussed in Ref.[4], this approximation can be invoked only in the representation that diagonalizes the effective system's Hamiltonian $H^{eff} = H_0 + V - (i/2)\Gamma_1 |1><1|$. The procedure therefore includes transformation to the representation which diagonalizes this Hamiltonian, following the Redfield procedure in this representation then transforming back to the representation defined in terms of states $|0>$ and $|1>$. It yields[4]

$$\dot{\sigma}_{n,n'} = 0 = -i[H_0 + V, \sigma]'_{n,n'} - \frac{1}{2}\Gamma_1(\delta_{n,1} + \delta_{n',1})\sigma_{n,n'} + \sum_{n_1=0}^{1}\sum_{n_2=0}^{1} R_{n,n',n_1,n_2} \sigma_{n_1,n_2} \quad ; \quad n,n' = 0,1$$
$$(82)$$

where the prime on the commutator denotes that it has been modified by eliminating the terms incompatible with a steady-state driven by state $|0>$ as discussed above, and where the tetradic elements $R_{n_1,n_2,n_3,n_4}$ may be expressed in terms of the correlation function $<F_{11}(t)F_{11}(0)>$. Solving (82) for the steady state defined by a fixed $\rho_{0,0}$ finally yields the steady-state values of $\sigma_{0,1}$, $\sigma_{1,0}$ and $\sigma_{1,1}$; see Appendix C for more details.

Having obtained explicit expressions for $\sigma_{0,1}$ and $\sigma_{11}$, Eqs. (77) and (78) can be used to obtain $\sigma_{1,r}$ and $\sigma_{rr}$. From these we get the energy resolved flux (or steady state rate) $k_{0 \to r} = \eta \sigma_{r,r} / \sigma_{0,0}$ and the total flux $k_{0 \to R} = \sum_r k_{0 \to r}$ into the right manifold. Numerical results for these observables are shown below. The analytical expressions are very cumbersome, but can be simplified in the limit where the energy gap, $E_1$-$E_0$ is much larger than all other energy parameters in the system, i.e., $E_1$-$E_0 \gg |V_{1,0}|, \kappa, \Gamma_1$ ($\kappa$ is defined in Eq. (63)). In this limit we get

$$k_{0 \to r} = \frac{|V_{0,1}|^2}{(E_1 - E_0)^2 + (\Gamma_1/2)^2} \left[ 2\pi \delta(E_0 - E_r) |V_{1,r}|^2 + \frac{|V_{1,r}|^2 \kappa e^{-\beta(E_1 - E_0)}}{(E_1 - E_r)^2 + (\Gamma_1/2)^2} \right] \quad (83)$$

and

$$k_{0 \to R} = \frac{|V_{0,1}|^2 \Gamma_{1R}}{(E_1 - E_0)^2 + (\Gamma_1/2)^2} \left[ 1 + \frac{\kappa}{\Gamma_1} e^{-\beta(E_1 - E_0)} \right] \quad (84)$$

We can also repeat the procedure of Eqs. (29)-(32) to find an expression for the transmission coefficient:

$$\mathcal{T}(E_0, E) = \mathcal{T}_{el}(E_0) \left[ \delta(E_0 - E) + \frac{(\kappa/2\pi) e^{-\beta(E_1 - E_0)}}{(E_1 - E)^2 + (\Gamma_1/2)^2} \right] \quad (85)$$

and

$$\mathcal{T}(E_0) = \int dE \mathcal{T}(E_0, E) = \mathcal{T}_{el}(E_0) \left[ 1 + \frac{\kappa}{\Gamma_1} e^{-\beta(E_1 - E_0)} \right] \quad (86)$$

where, as before (c.f. Eq. (32))

$$\mathcal{T}_{el}(E_0) = \frac{\Gamma_{1L} \Gamma_{1R}}{(E_1 - E_0)^2 + (\Gamma_1(E_0)/2)^2} \quad (87)$$

These results show clearly the coherent and the incoherent-activated components of the flux and the transmission. We note that similar results, but without the temperature dependent exponential, were obtained previously for the use of a similar model for resonance Raman scattering.[5] The erroneous absence of this term can be traced to the improper use of the Redfield approximation in a basis set that does not diagonalize the system's Hamiltonian, as explained above and in Reference [4]

Consider again the use of this system as a model for a molecular conductor bridging between two metal contacts. In the absence of thermal relaxation the conduction of the resulting junction is given by the Landauer formula (46). Here the issue is more complex because the transmitted electron can carry energy different from $E_0$. A generalization of Eq. (46) for the present situation can be obtained in the weak metal-bridge coupling limit where the current can be written in the form(17)

$$I = \frac{e}{\pi\hbar} \int_0^\infty dE_0 \int_0^\infty dE \mathcal{T}(E_0, E) \left[ f(E_0)(1 - f(E + e\Phi)) - f(E_0 + e\Phi)(1 - f(E)) \right] \qquad (88)$$

where $f(E)$ is the Fermi Dirac distribution and $\Phi$ is the potential drop between the right and left electrodes. For small bias and low enough temperature (so that $f(E + e\Phi) \sim f(E) - e\Phi\delta(E - E_F)$) this leads to

$$g(E_0) = \frac{I}{\Phi} = \frac{e^2}{\pi\hbar} \mathcal{T}_{el}(E_0) \left( 1 + (1 - f(E_1)) \frac{\kappa}{\Gamma_1} e^{-\beta(E_1 - E_0)} \right) \qquad (89)$$

While this result was obtained for a simple model of a single state bridge in the weak coupling limit, its structure is characteristic, displaying an elastic tunneling and thermally activated components.

## 4. Flux through an N-site bridge

The steady state density matrix formalism described in the previous section is easily formulated also for more general situations. As an example we outline here the generalization of the model discussed above (transmission through one intermediate state to a model for transmission through $N$ intermediate levels, Fig. 2c. We will continue to use the language corresponding to transport of non-interacting electrons through a simple molecular bridge connecting two simple metal electrodes. The metal electrodes are represented by the continuous manifolds of states $L \equiv \{|l>\}$ and $R \equiv \{|r>\}$. The molecular bridge now consists of N states, $\{|n>\}; n = 1, 2, ..., N$. The Hamiltonian is

$$H = H_M + H_B + F + H_C + H_{CM} \qquad (90)$$

where, as before $H_B + F$ represent the thermal environment and its coupling to the electronic system, $H_M$ is the bridge Hamiltonian

$$H_M = H_0 + V$$
$$H_0 = \sum_{n=1}^N E_n |n><n| \quad ; \quad V = \sum_{n=1}^{N-1} V_{n,n+1} |n><n+1| + V_{n+1,n} |n+1><n| \qquad (91)$$

$H_C$ is the Hamiltonian for the metal electrodes, our scattering continua

$$H_C = \sum_l E_l |l><l| + \sum_r E_r |r><r| \qquad (92)$$

and $H_{CM}$ is the electrode - molecule coupling

$$H_{CM} = \sum_l V_l + \sum_r V_r$$
$$V_l = V_{l,1}|l\rangle\langle 1| + V_{1,l}|1\rangle\langle l| \qquad V_r = V_{r,N}|r\rangle\langle N| + V_{N,r}|N\rangle\langle r| \qquad (93)$$

Note that in (91) and (93) we assume nearest neighbor coupling, and in particular the metal states are taken to be coupled only to the nearest molecular states, 1 and $N$. The molecule-thermal bath coupling is assumed to be of the form

$$F = \sum_{n=1}^{N} F_{n,n} |n\rangle\langle n| \qquad (94)$$

where the bath operators $F_{n,n}$ are characterized by their average and correlation functions. For the present model we take $<F_{n,n}> = 0$ and $<F_{n,n}(0)F_{n',n'}(t)> = C(t)\delta_{n,n'}$, where $<>$ denotes here equilibrium thermal average. As in Eq. (63) we sometimes use $C(t) = \kappa(2\tau_c)^{-1}\exp(-|t|/\tau_c)$.

Our problem is again to compute the steady-state flux into a right-continuum level $|r\rangle$, given that the system is driven by state $|0\rangle$, a representative state of the $L$ manifold. To this end we consider the Liouville equation

$$\dot\rho = -i[H_M + H_C + H_{CM}, \rho] - i[F, \rho] \qquad (95)$$

where, without changing notation, $F$ and $\rho$ now denote the transformed operators, $F(t) = \exp(iH_B t)F\exp(-iH_B t)$ and $\rho(t) = \exp(iH_B t)\rho\exp(-iH_B t)$. In (95) the first commutator corresponds to the deterministic part of the time evolution and the second – to the thermal part. In analogy with the development of Section 3, the deterministic part of the equations should be modified when applied to a steady state driven by $|0\rangle$.[12] In fact, this part is most straightforwardly derived from the amplitude equations (analogs of (19)-(22))

$$\dot c_0 = -iE_0 c_0 \qquad (96)$$

$$\dot c_1 = -iE_1 c_1 - iV_{1,0} c_0 - iV_{1,2} c_2 - i\sum_{l\neq 0} V_{1,l} c_l \qquad (97)$$

$$\dot c_n = -iE_n c_n - iV_{n,n-1} c_{n-1} - iV_{n,n+1} c_{n+1} \;;\; n=2,...,N-1 \qquad (98)$$

$$\dot c_N = -iE_N c_N - iV_{N,N-1} c_{N-1} - i\sum_r V_{N,r} c_r \qquad (99)$$

$$\dot c_l = -iE_l c_l - iV_{l,1} c_1 - (\eta/2) c_l \qquad (100)$$

$$\dot c_r = -iE_r c_r - iV_{r,N} c_N - (\eta/2) c_r \qquad (101)$$

using the procedure described in Appendix B. For the thermal part, we assume (as in Eqs. (71)-(75)) that it can be omitted from all equations for density matrix elements involving $\ell$ or $r$ states. The resulting equations for the time evolution of the density matrix are

$$\rho_{00} = \text{constant} \tag{102}$$

$$\dot{\rho}_{0,n} = 0 = -iE_{0,n}\rho_{0,n} + iV_{n-1,n}\rho_{0,n-1} + iV_{n+1,n}\rho_{0,n+1} \\ -\left((\Gamma_L/2)\delta_{n,1} + (\Gamma_R/2)\delta_{n,N}\right)\rho_{0,n} - i[F,\rho]_{0,n} \tag{103}$$

$$\dot{\rho}_{n,n'} = 0 = -iE_{n,n'}\rho_{n,n'} - i\left(V_{n,n-1}\rho_{n-1,n'} + V_{n,n+1}\rho_{n+1,n'} - \rho_{n,n'-1}V_{n'-1,n'} - \rho_{n,n'+1}V_{n'+1,n'}\right) \\ -(\Gamma_L/2)(\delta_{n,1} + \delta_{n',1})\rho_{n,n'} - (\Gamma_R/2)(\delta_{n,N} + \delta_{n',N})\rho_{n,n'} - i[F,\rho]_{n,n'} \tag{104}$$

$$\dot{\rho}_{0,r} = 0 = -iE_{0,r}\rho_{0,r} + iV_{N,r}\rho_{0,N} - (\eta/2)\rho_{0,r} \tag{105}$$

$$\dot{\rho}_{n,r} = 0 = -iE_{n,r}\rho_{n,r} - iV_{n,n+1}\rho_{n+1,r} - iV_{n,n-1}\rho_{n-1,r} \\ + iV_{N,r}\rho_{n,N} - \left((\Gamma_L/2)\delta_{n,1} + (\Gamma_R/2)\delta_{n,N}\right)\rho_{n,r} \tag{106}$$

$$\dot{\rho}_{r,r} = 0 = -iV_{r,N}\rho_{N,r} + iV_{N,r}\rho_{r,N} - \eta\rho_{r,r} \tag{107}$$

where $\Gamma_L = 2\pi\sum_l |V_{1,l}|^2 \delta(E_0 - E_l)$ and $\Gamma_R = 2\pi\sum_r |V_{N,r}|^2 \delta(E_0 - E_r)$. Associated with these are energy shifts of states 1 and $N$ that were absorbed into $E_1$ and $E_N$. As before we will disregard the energy dependence of these widths and shifts. Note that equations similar to (105) - (107) exist also for elements involving the $L$ manifold, however in what follows we focus on transmission into the $R$ manifold.

The following steps are identical to those taken to solve Eqs. (71)-(75). Again we note that these equations are grouped so that (102)-(104) describe a pumped (by the driving state $|0>$) and damped thermally relaxing N level system, while (105)-(107) do not depend on the interaction with the heat bath. Consider first Eqs. (102)-(104). They are solved by carrying out the same reduction procedure (transforming to a representation in which the effective system's Hamiltonian is diagonal, following the Redfield procedure modified for steady-state and transforming back to the local, site state, representation) as described in Section 3 and in Appendix C. The final result is an equation similar to (82)

$$\dot{\sigma}_{n,n'} = 0 = -i[H_0 + V, \sigma]_{n,n'} + \left[-\frac{1}{2}\Gamma_R(\delta_{n,N} + \delta_{n',N}) - \frac{1}{2}\Gamma_L(\delta_{n,1} + \delta_{n',1})\right]\sigma_{n,n'} + \sum_{n_1=0}^{N}\sum_{n_2=0}^{N} R_{n,n',n_1,n_2}\sigma_{n_1,n_2}$$
$$n,n' = 0...N \quad (n \times n' \neq 0) \tag{108}$$

where again the primed commutator is modified to satisfy the steady state restrictions[19] and where the tetradic elements $R_{n_1,n_2,n_3,n_4}$ are linear combinations of Fourier and Fourier-Laplace transforms of the correlation functions $<F_{n,n}(t)F_{n,n}(0)>$ (explicit expressions are given in Ref.[4]). Eqs. (108) (with $\sigma_{0,0}$=constant) constitute a set of linear equations for the elements $\sigma_{n,n'}$ of the reduced molecular density matrix, and in particular will yield explicit expressions for the elements $\sigma_{n,N}$, $n=0...N$ that are needed below.

Turning now to Eqs. (105)-(107), we note that they do not depend on the interaction with the heat bath, and can therefore be converted into equations for $\sigma$ by tracing over the bath. In particular, taking this trace in Eqs. (105) and (106) leads to equations for $\sigma$ elements that can be put into the form

$$\mathbf{A} \cdot \mathbf{y} = \mathbf{x} \tag{109}$$

where $\mathbf{x}$ and $\mathbf{y}$ are the vectors

$$\mathbf{y} = \begin{pmatrix} \sigma_{0,r} \\ \vdots \\ \sigma_{n,r} \\ \vdots \\ \sigma_{N,r} \end{pmatrix} \quad ; \quad \mathbf{x} = -iV_{N,r} \begin{pmatrix} \sigma_{0,N} \\ \vdots \\ \sigma_{n,N} \\ \vdots \\ \sigma_{N,N} \end{pmatrix} \tag{110}$$

and $\mathbf{A}$ is the matrix

$$\mathbf{A} = \begin{pmatrix} -iE_{0,r}-\eta/2 & 0^* & 0 & \cdots & \cdots & 0 & 0 \\ -iV_{1,0} & -iE_{1,r}-\Gamma_L/2 & -iV_{1,2} & 0 & \cdots & & \\ 0 & & \ddots & & & & \\ \vdots & & -iV_{n,n-1} & -iE_{n,r}-\eta/2 & -iV_{n,n+1} & \vdots & \vdots \\ \vdots & & & \ddots & & 0 & 0 \\ & & 0 & & -iV_{N-1,N-2} & -iE_{N-1,r}-\eta/2 & -iV_{N-1,N} \\ 0 & & \cdots & & 0 & -iV_{N,N-1} & -iE_{N,r}-\Gamma_R/2 \end{pmatrix} \tag{111}$$

In (111) the element marked $0^*$ is a zero that replaces a term $-iV_{0,1}$ in the original Liouville equation[12] (see also discussion below Eq. (59)). Solving for $\mathbf{y}$ and using (107) in the form $\eta\sigma_{rr} = 2\,\text{Im}(V_{r,N}\sigma_{N,r})$ yields an expression for $\eta\sigma_{r,r}$ in terms of the elements $\sigma_{n,N}$, $n=0,1,...,N$, that were obtained before. This provides a straightforward numerical procedure for evaluating $\eta\sigma_{r,r}$, i.e., the energy resolved flux transmitted into the right manifold, i.e. into the right electrode.

Examples of results predicted by this model (see Fig. 2c) for transmission through a thermally relaxing bridge are shown in Figures 3-5. Fig. 3 depicts the energy resolved transmission probability, $T(E_0,E)$ for electrons incident with energy $E_0$ at several temperatures. The following model parameters were used: $N=3$, $\Delta E=E_{n,0}=E_n-E_0=3000\text{cm}^{-1}$, ($n=1...3$), $V_{n,n+1}=200\text{cm}^{-1}$, $\Gamma_L=\Gamma_R=160\text{cm}^1$ $\tau_c=0$ and $\kappa=10\text{cm}^{-1}$. The transmitted flux plotted against $E-E_0$ is seen to consist of two components: elastic tunneling at energy $E=E_0$ and activated tunneling in an energy range corresponding to the bridge states. To avoid numerical problems, the displayed results were obtained at finite resolution by using $\eta=10\text{cm}^{-1}$.

Obviously, the tunneling and activated components seen in Fig. 4 should depend differently on system parameters. To see this we have used the corresponding quantities $T_t$ and $T_a$ obtained numerically as integrals over the corresponding peaks in Fig. 3. Figure 4 shows the dependence of these components as well as the overall transmission probability $T(E_0)=T_t(E_0)+T_a(E_0)$ on temperature, using the same system's parameters as above. Figure 5 shows their behavior as functions of the bridge length $N$. It is seen that the tunneling component is temperature independent and decreases exponentially with increasing bridge length, while the activated component does not depend (in the range displayed) on the bridge length, and depends exponentially on the inverse temperature. The overall transmission probability shows the characteristic temperature and bridge length dependence already studied in Ref. [4]. In particular the apparent insensitivity of the activated component to bridge length holds only at the intermediate length regime, and actually reflects a dependence on N of the form $(\alpha_1+\alpha_2 N)^{-1}$ with $\alpha_1 \gg \alpha_2$ [4]. It is also important to note that the insensitivity of the tunneling component to temperature is a property of the present model, appropriate for the *weak thermal coupling case*. (This roughly corresponds to the situation where either the energy gap, $|E_{0,1}|$, or the 'bandwidth' characterized by $V_{n,n+1}$ are much larger than level broadening due to thermal interactions). In the opposite limit we find[6] that destruction of coherence due to thermal interactions affects the tunneling probability in a way that depends on temperature.

For fixed system parameters, the relative importance of the coherent and incoherent transmission components changes with the distance from resonance. Both time and energy scale considerations suggest that as the energy gap $\Delta E$ becomes smaller the relative importance of the incoherent component increases. This is indeed seen in Fig. 6 that shows the relative magnitudes of these components as functions of $\Delta E$. It should be emphasized that close to resonance it is no longer possible to represent the overall transmission as an additive combinations of coherent and incoherent contributions.

## 5. Summary and conclusions

In this paper we have developed frameworks for the description of steady states of open quantum mechanical systems. In the absence of thermal relaxation the developed formalism provides a reformulation of standard time independent scattering theory that is more flexible in the way in which the state that drives the system (the equivalent of the incoming wave in the standard formalism) is defined. When thermal interactions are included in the target model the theory yields a description of the scattering process in Liouville space yielding a scattered flux that includes a thermal incoherent component. We have used this approach in conjunction with the Redfield approximation to study tunneling through a metal-molecule-metal junction, including thermal relaxation and dephasing in the molecular component. We have found that zero bias conduction through such junctions involves both tunneling and activated components. This leads to a generalization of the Landauer formula of conduction to situations involving thermal interactions. The coherent tunneling and the incoherent activated components depend differently on the temperature, the barrier height and the molecular chain length.

These results capture the essential phenomenology of molecular conduction in the linear (ohmic) regime in the presence of thermal interactions. We should keep in mind that the simplifications used in constructing and analyzing Eqs. (102)-(107) (or (71)-(75)). These were the neglect of thermal interactions in all equations for density matrix elements involving states of the continuous manifolds (see paragraph below Eq. (70)) and the use of the Redfield approximation that limits the validity of our result to the weak thermal coupling limit. In particular, as noted in the previous section, the apparent insensitivity of the coherent part of the transmission to the thermal interactions holds (approximately) only in this limit. These issues will be discussed further in a subsequent publication.

**Appendix A**

It is important to keep in mind that a "steady state" in an open system is not unique and depends on the choice of the boundary system. (In the general case of non-linear equations even this does not guarantee uniqueness, but in our quantum mechanical applications the equations are linear so this will not be a point of concern). One should choose these boundary conditions so as to correspond to the physical realization that we want to describe. Suppose for example that Eq. (1), written explicitly as $\dot{C}_j = F_j(\{C_j\})$, represents a master equation that describes a transition from an initial state 0 to a final state $N+1$ through a state of intermediate states $n=1,2,...N$. The variables $\{C_j\}$, $j=0,1,...N+1$ may represent densities or probabilities. The $N$ steady state equations $F_j(\{C_j\};C_0) = 0$, $j = 1,...,N$, obtained by replacing the equation for $\dot{C}_0$ by the boundary condition $C_0 =$ constant and the equation for $\dot{C}_{N+1}$ by $C_{N+1} = 0$, can be used to obtain the corresponding steady state values of $C_1,...,C_N$. The steady state flux, $J_{ss}$, can then be obtained as the rate at which population flows into (and out of) $C_{N+1}$ and the corresponding steady state rate is $J_{ss}/C_0$. This or an equivalent procedure is often used to evaluate asymptotic (i.e. long time) decay rates of transient processes, provided that conditions for an early formation of a quasi steady-state situation are satisfied. (See Ref. [4] for a more detailed discussion of this point).

In the quantum mechanical examples discussed in Sect. 2 an analogous approach is taken, however it differs from the procedure just described in two important aspects. First, the variables $C$ are quantum mechanical amplitudes, and do not describe completely the initial state. Therefore, the 'boundary condition' $C_0$=constant is supplemented by the term $\exp(-iE_0 t)$, see e.g. Eq. (8), that set the energy scale of the process studied. Second, while in many applications a continuum of states represent a bath whose state is not specified explicitly, in other situations a knowledge of the flux into a specific final state of a continuum is needed; for example in an experiment that monitors the final energy of an emitted photon or a scattered free particle. In the latter case each energy state of the continuum should be considered as a part of the system (in the classical analogy - as part of the $N$-state system discussed above) and the sink is introduced artificially by adding a small imaginary part to the corresponding energy, see e.g. Eq. (9). (The details of this imaginary addition do not affect the final result; it just serves to set the correct directionality of the process in much the same as a similar term is used in the Green's function of scattering theory).

Again, it should be emphasized that one could investigate in principle other steady states, for example a process where $|C_0|$ is given but the energy is not or processes where other $C_j$'s are restricted in some ways - if a relevant physical case could be identified. The particular procedure used in Section 2 is set so that state 0 plays the part of an incoming state, while $\{j\}$ is a manifold of outgoing states. Indeed we find in Section 2 that this approach reproduces basic results of scattering theory.

**Appendix B**

Here we show that the density matrix equations (54)-(59), or Eqs. (71)-(75) without the terms involving $F$ lead to Eq. (76). First we show that Eqs. (71)-(75) are consistent with the amplitude equations. The latter are (cf. Eqs. (19)-(22) with $\{j\}$ standing for both $\{r\}$ and $\{\ell\}$ manifolds)

$$\dot{c}_0 = -iE_0 c_0 \tag{112}$$

$$\dot{c}_1 = -iE_1 c_1 - iV_{1,0} c_0 - i\sum_j V_{1,j} c_j \tag{113}$$

$$\dot{c}_j = -iE_j c_j - iV_{j,1} c_1 - (\eta/2) c_j \tag{114}$$

At steady state (Eq.(23)) Eq. (114) becomes

$$i(E_0 - E_j) c_j - iV_{j,1} c_1 - (\eta/2) c_j = 0 \tag{115}$$

Inserting $c_j$ from (115) into (113) and using Eq. (24) yields

$$\dot{c}_1 = -i\tilde{E}_1 c_1 - iV_{1,0} c_0 - (1/2)\Gamma_1 c_1 \tag{116}$$

where $\Gamma_1 = \Gamma_1(E_0) = 2\pi \sum_j |V_{1j}|^2 \delta(E_0 - E_j) = \Gamma_{1L}(E_0) + \Gamma_{1R}(E_0)$ and where $\tilde{E}_1 = E_1 + \Lambda_1(E_0)$ and $\Lambda_1(E_0) = PP\sum_j |V_{1j}|^2 /(E_0 - E_j) = \Lambda_{1L}(E_0) + \Lambda_{1R}(E_0)$. (*PP* stands for *Principal Part*). Using $\rho_{i,j} = c_i c_j^*$ and focusing on steady state this yields

$$\dot{\rho}_{0,1} = 0 = -i(E_0 - \tilde{E}_1)\rho_{0,1} + iV_{0,1}\rho_{0,0} - \frac{1}{2}\Gamma_1 \rho_{0,1} \tag{117}$$

$$\dot{\rho}_{1,1} = 0 = -iV_{1,0}\rho_{0,1} + iV_{0,1}\rho_{1,0} - \Gamma_1 \rho_{11} \tag{118}$$

$$\dot{\rho}_{0,j} = 0 = -i(E_0 - E_j)\rho_{0,j} + iV_{1,j}\rho_{0,1} - (\eta/2)\rho_{0,j} \tag{119}$$

$$\dot{\rho}_{1,j} = 0 = -i(\tilde{E}_1 - E_j)\rho_{1,j} - iV_{1,0}\rho_{0,j} + iV_{1,j}\rho_{1,1} - \frac{1}{2}\Gamma_1 \rho_{1,j} \tag{120}$$

$$\dot{\rho}_{j',j} = 0 = -i(E_{j'} - E_j)\rho_{j',j} - iV_{j',1}\rho_{1,j} + iV_{1,j}\rho_{j',1} - \eta\rho_{j',j} \tag{121}$$

Next we use Eqs. (117)-(121) to derive Eqs. (27) and hence (28). Using (121) with $j=j'$ yields

$$\eta\rho_{jj} = 2\text{Im}(V_{j1}\rho_{1j}) \tag{122}$$

Also, Eqs. (118)-(121) lead to (with $\tilde{E}_{0,1} = E_0 - \tilde{E}_1$)

$$\rho_{0,1} = \frac{V_{0,1}\rho_{0,0}}{\tilde{E}_{0,1} - (1/2)i\Gamma_1(E_0)} \tag{123}$$

$$\rho_{0,j} = \frac{V_{0,1}V_{1,j}\rho_{0,0}}{(\tilde{E}_{0,1} - (1/2)i\Gamma_1(E_0))(E_{0,j} - (1/2)i\eta)} \tag{124}$$

$$\rho_{1,1} = \frac{|V_{0,1}|^2 \rho_{0,0}}{\tilde{E}_{0,1}^2 + ((1/2)\Gamma_1(E_0))^2} \tag{125}$$

We note in passing that comparing Eqs. (118) and (57) suggests the following identity

$$-i\sum_j V_{1,j}\rho_{j,1} + i\sum_j V_{j,1}\rho_{1,j} = -\Gamma_1(E_0)\rho_{11} \tag{126}$$

Eqs. (124) and (125) yield after some algebra

$$\rho_{1,j} = \frac{|V_{0,1}|^2 V_{1,j}\rho_{0,0}}{[\tilde{E}_{0,1}^2 + ((1/2)\Gamma_1(E_0))^2](E_{0,j} - (1/2)i\eta)} \tag{127}$$

Using (127) in (122) leads to Eq. (27).

**Appendix C**

Here we outline the procedure used to obtain system's density matrix elements from Eqs. (71)-(72) within the Redfield approximation. First we find the transformation that diagonalizes $H^{eff} = H_0 + V - (i/2)\Gamma_1 |1><1|$. In the new representation, defined in terms of eigenstates $|a>$ and $|b>$ (with corresponding eigenvalues $E_a-(1/2)i\Gamma_a$ and $E_b-(1/2)i\Gamma_b$) the overall system-bath Hamiltonian is $H^{eff} + H_B + F$, where

$$H^{eff} = (E_a\text{-}i\Gamma_a/2)|a><a| + (E_b\text{-}i\Gamma_b/2)|b><b| \tag{128}$$

and

$$F = F_{a,a}|a><a| + F_{b,b}|b><b| + F_{a,b}|a><b| + F_{b,a}|b><a| \tag{129}$$

with

$$F_{a,a} = \left[\frac{1}{2} - \frac{|\Delta|}{2\tilde{E}}\right]F_{1,1} \; ; \quad F_{b,b} = \left[\frac{1}{2} + \frac{|\Delta|}{2\tilde{E}}\right]F_{1,1} \; ; \quad F_{a,b} = F_{b,a} = \frac{-|V_{0,1}|}{2\tilde{E}}F_{1,1} \quad (130)$$

and

$$\Delta = (E_1 - E_0 - i\Gamma_1/2)/2 ; \quad \tilde{E} = \sqrt{|V_{0,1}|^2 + |\Delta|^2} \; . \quad (131)$$

In this representation the Liouville equation $\dot{\rho} = -i[H^{eff}, \rho] - i[F, \rho]$ is reduced to four equations for the system's density matrix $\sigma'_{n,n'}$ ($n,n'=a,b$) using a variation of the Redfield formalism[4] (we use $\sigma'$ to denote this density matrix in the diagonal basis). The difference from the standard procedure[16] lies in the fact that in the Redfield theory $\sigma_{j,k}(t)\exp(-iE_{j,k}t)$ with $E_{j,k} = E_j - E_k$ is assumed 'slow', while, as discussed above, at steady state $\sigma_{j,k}$ does not depend on time, so such a transformation is not needed. We get

$$\dot{\sigma}'_{a,a} = -A K \sigma'_{a,a} + A \sigma'_{b,b} - (\sqrt{AD} - \sqrt{AC}K)(\sigma'_{a,b} + \sigma'_{b,a}) - \Gamma_a \sigma'_{a,a}$$
$$\dot{\sigma}'_{b,b} = A K \sigma'_{a,a} - A \sigma'_{b,b} + (\sqrt{AD} - \sqrt{AC}K)(\sigma'_{a,b} + \sigma'_{b,a}) - \Gamma_b \sigma'_{b,b}$$
$$\dot{\sigma}'_{a,b} = \sqrt{AB}K\sigma'_{a,a} - \sqrt{AB}\sigma'_{b,b} + A \sigma'_{b,a} - iE_{a,b}\sigma'_{a,b} - (2\sqrt{BC}K + A - 2\sqrt{DB})\sigma'_{a,b} - (\Gamma_a + \Gamma_b)/2\sigma'_{a,b}$$
$$\dot{\sigma}'_{b,a} = \sqrt{AB}K\sigma'_{a,a} - \sqrt{AB}\sigma'_{b,b} + A \sigma'_{a,b} - iE_{b,a}\sigma'_{b,a} - (2\sqrt{BC}K + A - 2\sqrt{DB})\sigma'_{b,a} - (\Gamma_a + \Gamma_b)/2\sigma'_{b,a}$$

$$(132)$$

Where

$$A = \frac{|V_{0,1}|^2}{4\tilde{E}^2}\kappa ; \quad B = \frac{|\Delta|^2}{4\tilde{E}^2}\kappa ; \quad C = \left(\frac{1}{2} + \frac{|\Delta|}{2\tilde{E}}\right)^2 \frac{\kappa}{4} ; \quad D = \left(\frac{1}{2} - \frac{|\Delta|}{2\tilde{E}}\right)^2 \frac{\kappa}{4} \quad (133)$$

and where, in the Markovian ($\tau_c \to 0$) limit of Eq. (63)

$$\kappa = \int_{-\infty}^{\infty} e^{-iE_{a,b}\tau} <F_{1,1}(0)F_{1,1}(\tau)> d\tau \; ; \quad E_{a,b} = E_a - E_b \quad (134)$$

Also in Eq. (132) $K = e^{-\beta E_{b,a}}$ with $\beta = 1/k_B T$.

Eq. (132) is a set of linear differential equations for elements of $\sigma'$. Transforming back to the representation spanned by states $|0>$ and $|1>$ we get

$$\dot{\sigma}_{n,n'} = 0 = -i[H_0 + V, \sigma]'_{n,n'} - \frac{1}{2}\Gamma_1(\delta_{n,1} + \delta_{n',1})\sigma_{n,n'} + \sum_{n_1=0}^{1}\sum_{n_2=0}^{1} R_{n,n',n_1,n_2}\sigma_{n_1,n_2} \; ; \quad n,n'=0,1 \quad (135)$$

where the prime on the commutator denotes that it has been modified by eliminating the terms incompatible with a steady state driven by state $|0>$; see discussion below Eqs. (54)-(59). The explicit expressions for the **R** elements are quite cumbersome, but are easily calculated for any choice of model parameters. The desired steady state solution is obtained by solving the set of equations (135) (excluding $n=n'=0$) for a constant $\sigma_{0,0}$

Again, explicit general results for $\sigma$ are cumbersome but numerical results are easily computed. Simple expressions can be obtained in the limit where the energy gap is larger than all the other parameters, i.e. $E_1$-$E_0 \gg |V_{1,0}|, \kappa, \Gamma_1$. We get

$$\sigma_{1,1} = \frac{|V_{1,0}|^2 \sigma_{0,0}}{E_{0,1}^2 + (\Gamma_1/2)^2} + \frac{AK}{\Gamma_1}\sigma_{0,0} \tag{136}$$

$$\sigma_{0,1} = \frac{V_{0,1}\sigma_{0,0}}{E_{0,1} - i\Gamma_1/2} \tag{137}$$

In the same limiting case

$$A \equiv \frac{|V_{0,1}|^2}{4\tilde{E}^2}\kappa \cong \frac{|V_{0,1}|^2 \kappa}{E_{0,1}^2 + (\Gamma_1/2)^2} \tag{138}$$

so that

$$\sigma_{1,1} = \frac{|V_{0,1}|^2 \left(1 + \kappa K/\Gamma_1\right)}{E_{0,1}^2 + (\Gamma_1/2)^2}\sigma_{0,0} \tag{139}$$

Inserting (136) and (139) into (78) and (77) and taking the limit $\eta \to 0$ leads to Eqs. (83) and (84). To obtain (84) we use

$$\sum_r \frac{|V_{1,r}|^2}{E^2_{1,r} + (\Gamma_1/2)^2} \cong \int_{-\infty}^{\infty} \frac{|V_{1,r}|^2}{E^2_{1,r} + (\Gamma_1/2)^2}\rho_R(E_r)dE_r = \frac{2\pi |V_{1,r}|^2 \rho_R(E_1)}{\Gamma_1} = \frac{\Gamma_{1R}}{\Gamma_1} \tag{140}$$

Note that (84) could also be obtained by using Eqs. (70) and (126) to get

$$\eta\sum_r \sigma_{r,r} = \sum_r 2\,\text{Im}(V_{r,1}\sigma_{1,r}) = \Gamma_{1R}\sigma_{1,1} \tag{141}$$

and then using Eq. (139) for $\sigma_{1,1}$. We were able to carry this procedure also to the next order in the small parameter $|V_{1,0}|^2/E_{0,1}^2$. The results are

$$J_r = \eta\sigma_{r,r} =$$
$$\frac{|V_{0,1}|^2 \sigma_{0,0}}{E_{0,1}^2 + (\Gamma_1/2)^2}\left[2\pi\delta(E_0 - E_r)|V_{1,r}|^2 + \frac{|V_{1,r}|^2 \kappa e^{-\beta E_{1,0}}}{E^2_{1,r} + (\Gamma_1/2)^2}\left(1 + \frac{4|V_{0,1}|^2}{E_{0,1}^2 + (\Gamma_1/2)^2}\right)\right] \tag{142}$$

for the energy resolved flux, and

$$\sum_r \eta\sigma_{r,r} = \frac{|V_{0,1}|^2 \sigma_{0,0}\Gamma_{1R}}{E_{0,1}^2 + (\Gamma_1/2)^2}\left[1 + \frac{\kappa}{\Gamma_1}e^{-\beta E_{1,0}}\left(1 + \frac{4|V_{0,1}|^2}{E_{0,1}^2 + (\Gamma_1/2)^2}\right)\right] \tag{143}$$

for the total flux.

**Acknowledgements**   This research was supported in part by the United States-Israel Binational Science Foundation.

the final analysis will yield a term $|V_{0,1}|^2$ multiplied by the density $\rho_L$ of initial states in manifold $L$. This product is of order $\Omega^0$. The second equation corresponds to the backward flux from state n back into state 0. In the final analysis the specific term written above has to contain $V_{0,1}$ to higher order, that will vanish (even when multiplied by $\rho_L$ when $\Omega \rightarrow \infty$.

(13) The choice of a harmonic thermal bath with $F = \sum_\nu \lambda_\nu q_\nu$ where the sum is over the harmonic bath modes with coordinates $q_\nu$ and coupling strengths $\lambda_\nu$ corresponds to the standard model where the equilibrium positions of these modes are linearly shifted when the tunneling system is in the intermediate state |1>. However on the level of our treatment (i.e. in the Redfield approximation) an explicit form of $F$ is not needed, since only correlation functions such as $<F_{11}(t)F_{11}(0)>$ enter the reduced equations of motion.

(14) AG Redfield: IBM J. Res. Develop 1 (1957) 19.

(15) AG Redfield: Adv. Magen. Reson. 1 (1965) 1.

(16) WT Pollard, AK Felts, RA Friesner: Adv. Chem. Phys. 93 (1996) 77.

(17) Note that for the case under discussion $\mathcal{T}(E_0, E)$ does not depend on the directionality (left-to-right or right-to-left). It has been argued (see (18)) that simple expressions based on the Pauli principle such as Eq. (88) are not valid in the presence of inelastic processes including thermal relaxation. It seems that it may still be used in the weak metal-bridge coupling limit where the transmission process can be described as taking place between states localized on the two metals.

(18) S Datta: Electric transport in Mesoscopic Systems, Cambridge University Press, Cambridge, 1995.

(19) The terms corresponding to the primed commutator may be derived from Eqs. (96)-(99) by replacing $\rho$ by $\sigma$ and by using $\sigma_{n,n'} = c_n c_{n'}^*$.

**Figure Captions**

Fig. 1. A standard model for the decay of a prepared state coupled to a continuum

Fig. 2. Models for resonance scattering: (a) A double barrier structure with a quasi bound level in the intermediate well. (b) A standard approximation for resonance scattering from the potential (a), taking only the quasi-bound level in the well into account. The free particle states on the two sides of the barrier are depicted as continuous manifolds. (c) Same as (b), for a multi-well structure.

Fig. 3. Energy resolved transmission through a molecular bridge represented by the model of Fig 2c (N=3), supplemented by coupling to a heat bath as described in the text. The incident energy is $E_0$ and the transmission is depicted as a function of the transmitted energy for several temperatures. The curves showing increasing inelastic transmission near the bridge energy ($E$-$E_0 \sim 3000 cm^{-1}$) corresponds to the temperatures T=0, 300, 400, 500K, respectively. See text for the other parameters used in this calculation.

Fig. 4. The integrated elastic (dotted line) and activated (dashed line) components of the transmission, and the total transmission probability (full line) displayed as function of inverse temperature. Parameters are as in Fig. 3.

Fig. 5. The integrated elastic (dotted line) and inelastic (dashed line) components of the transmission, and the total transmission probability (full line) displayed as function of bridge length. Parameters are as in Fig. 3.

Fig. 6. Left panels: the integrated elastic (dotted line) and inelastic (dashed line) components of the transmission, and the total transmission probability (full line) displayed as functions of the distance $\Delta E$ from resonance. Right panels: The ratio R=$\mathcal{T}_a/\mathcal{T}$ between the activated component and the total transmission showing that far from resonance elastic transmission dominates. T=300K. Parameters are as in Fig. 3 except that the thermal coupling κ is $10 cm^{-1}$ (as in Fig. 2) in the upper panels and is $100 cm^{-1}$ in the bottom panels.

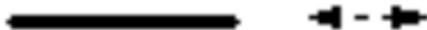
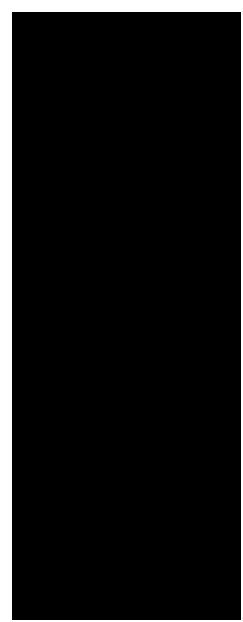

FIG. 1.

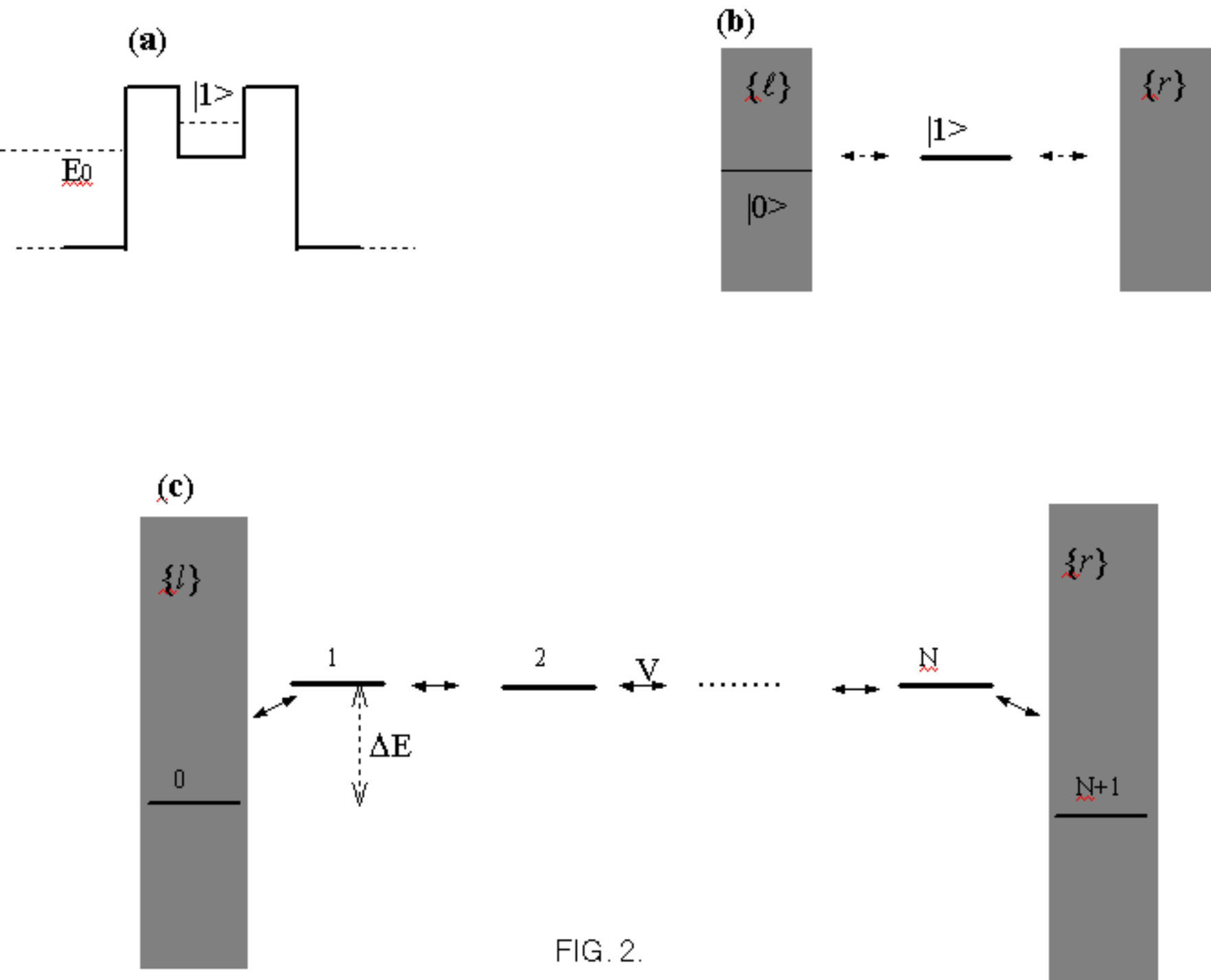

FIG. 2.

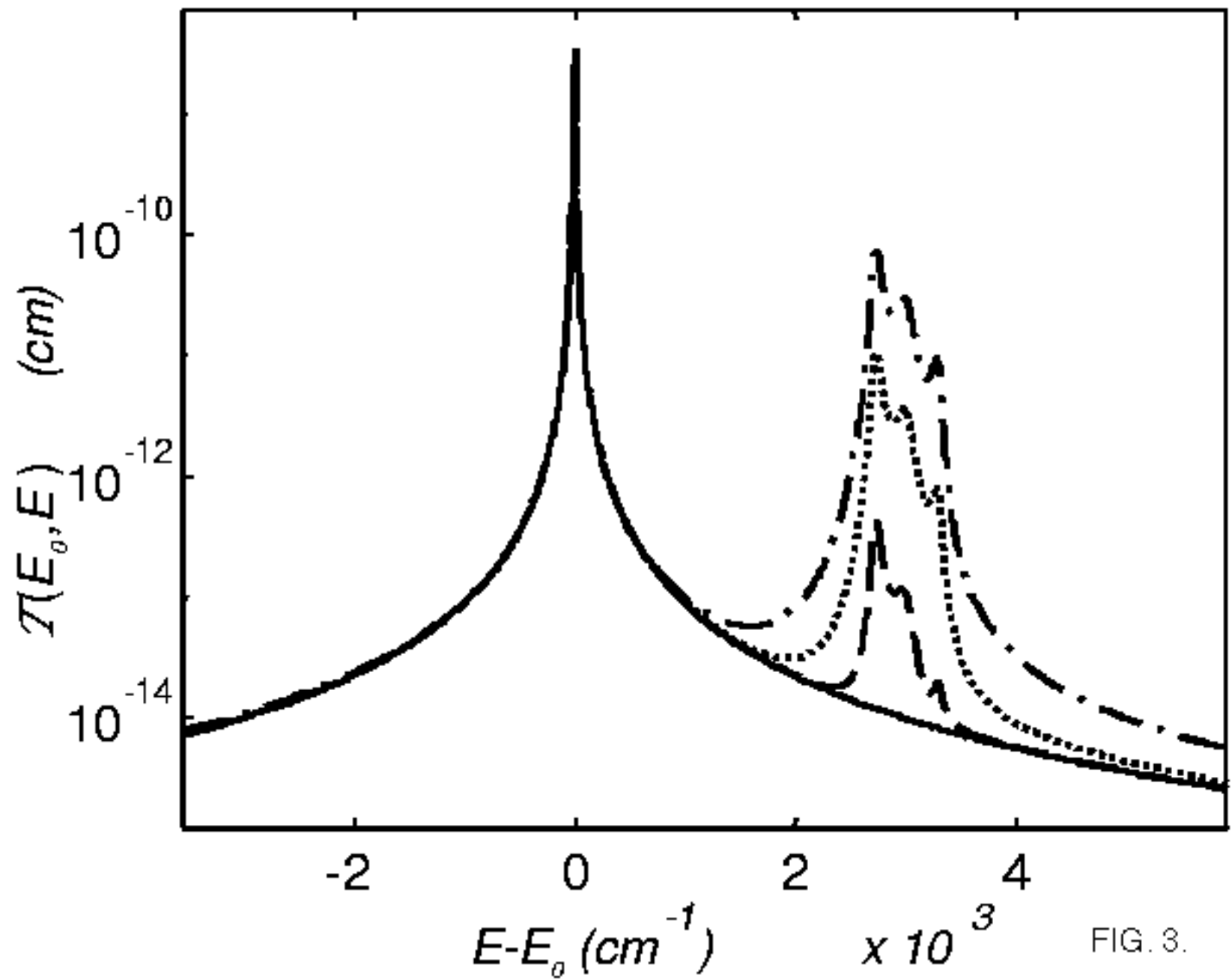

FIG. 3.

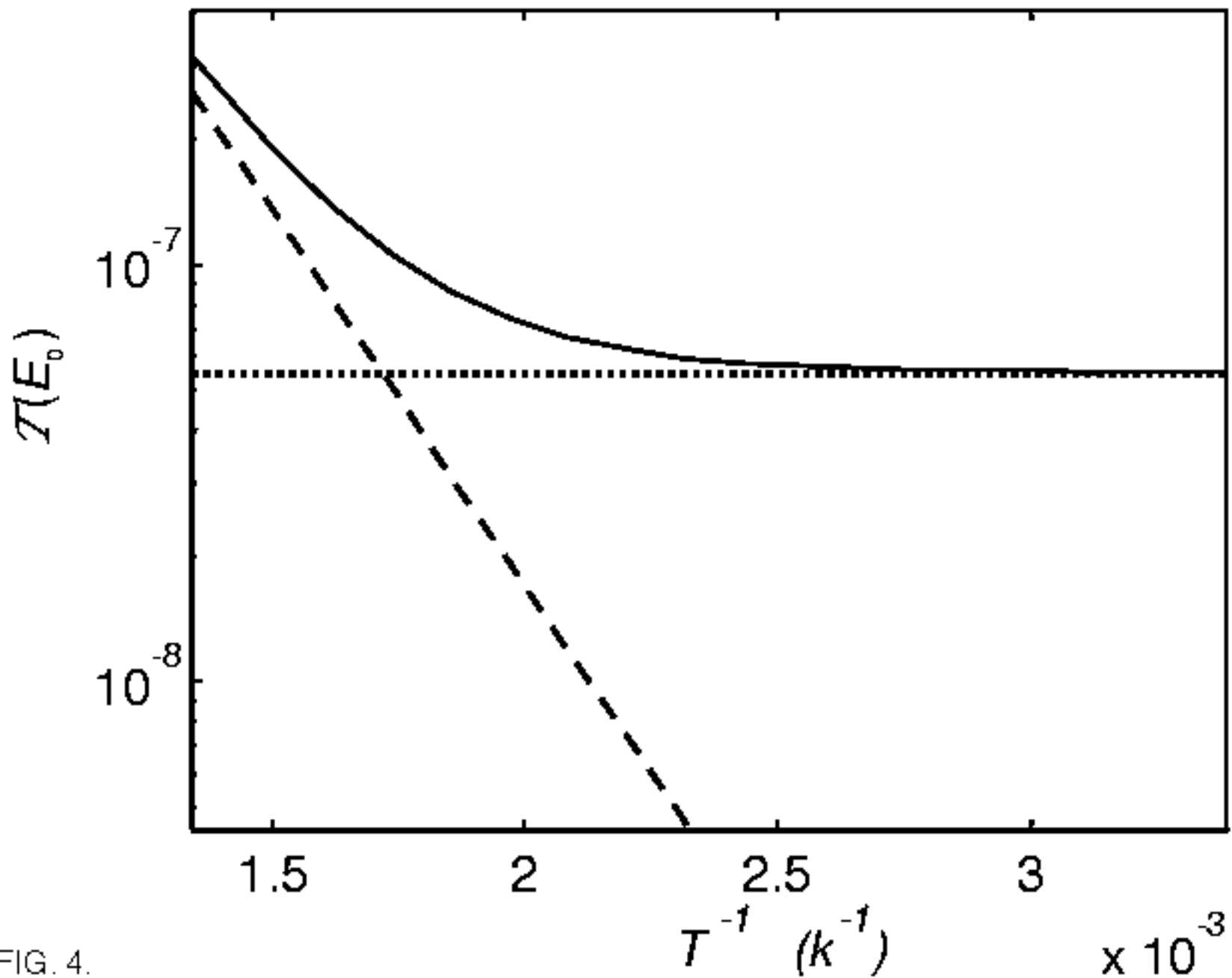

FIG. 4.

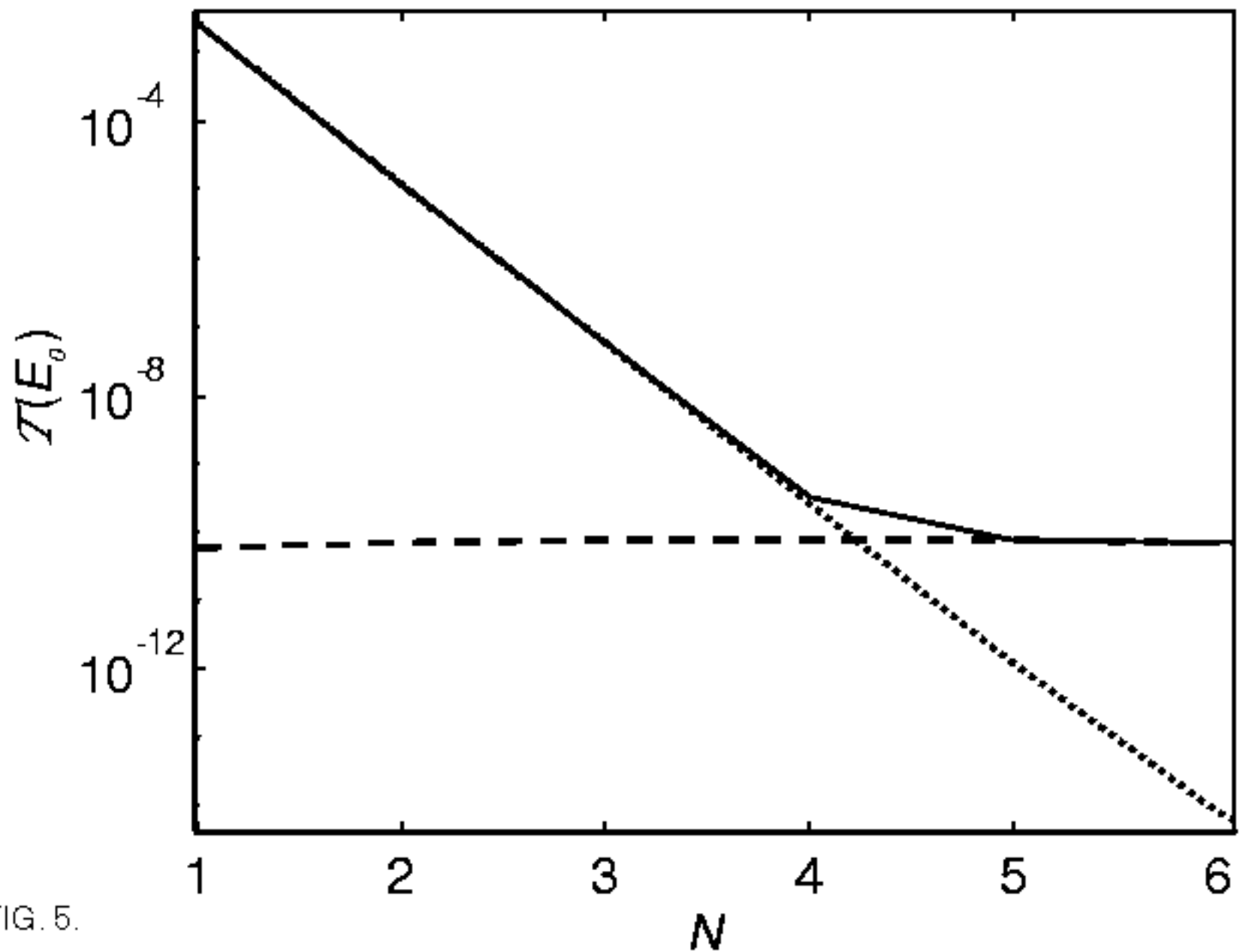

FIG. 5.

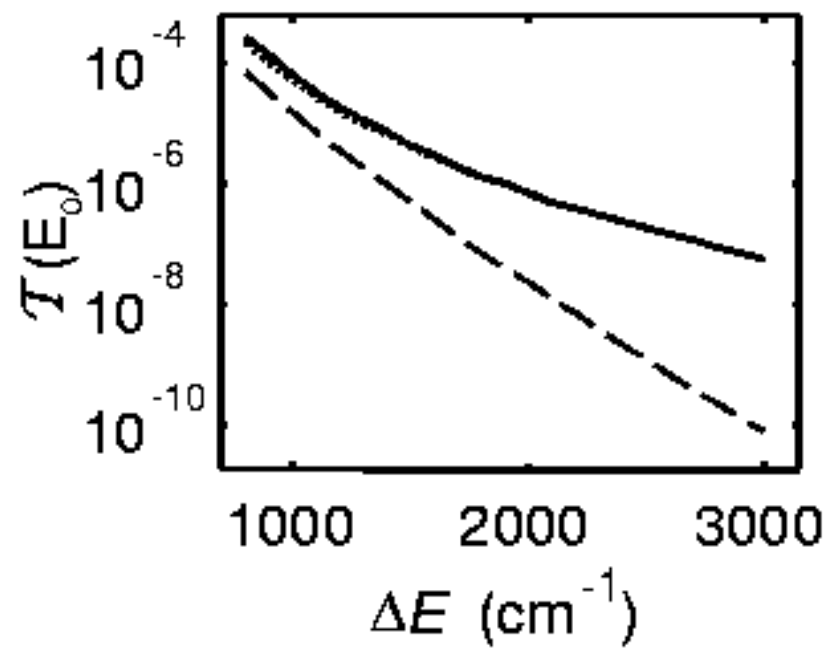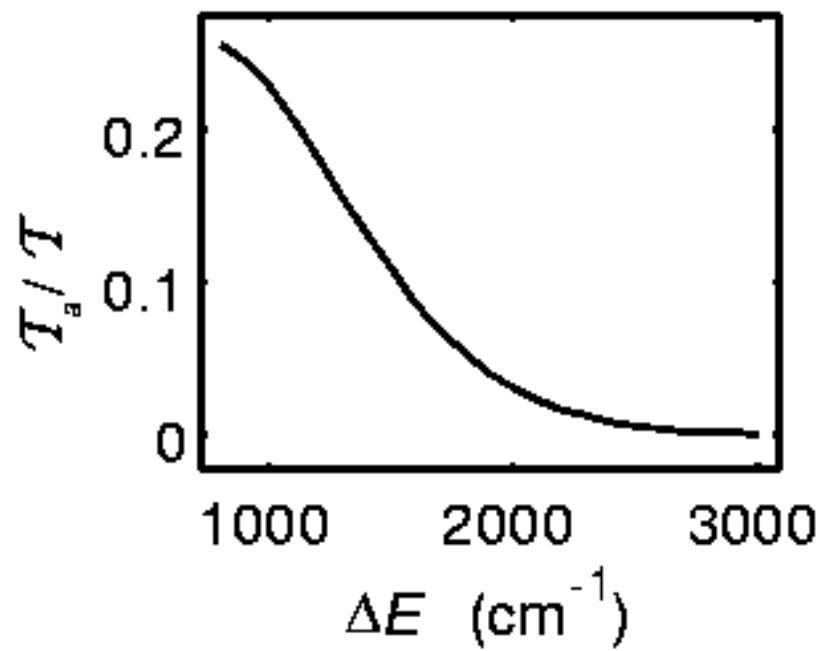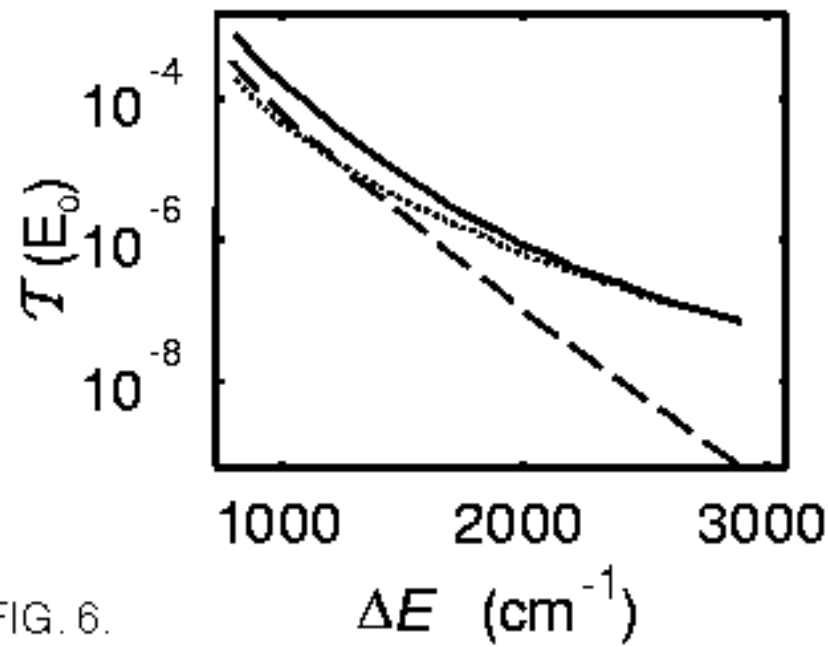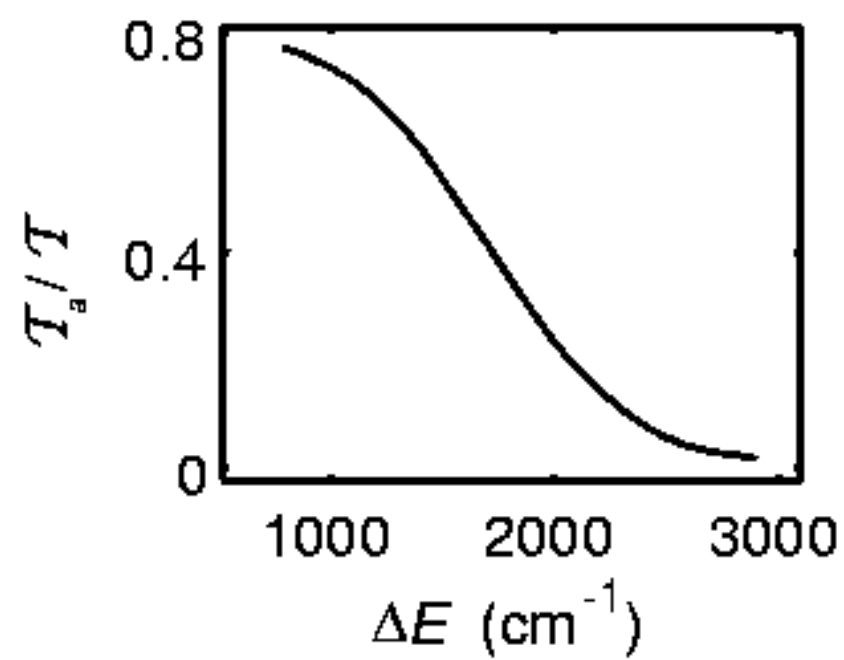

FIG. 6.